\documentclass[11pt,a4paper]{article}
\pdfoutput=1
\usepackage{jheppub}
\usepackage[sort&compress]{natbib}
\usepackage{url}
\usepackage{hyperref}
\usepackage{amsfonts}
\usepackage{epsfig}\usepackage{dcolumn}
\usepackage{bm}
\usepackage{slashed}
\graphicspath{{./figuras/}}
\usepackage{subfigure}
\usepackage{color}
\usepackage{mathrsfs}

\usepackage{listings}
\usepackage{color}
 
\definecolor{codegreen}{rgb}{0,0.6,0}
\definecolor{codegray}{rgb}{0.5,0.5,0.5}
\definecolor{codepurple}{rgb}{0.58,0,0.82}
\definecolor{backcolour}{rgb}{0.95,0.95,0.92}
 
\lstdefinestyle{mystyle}{
    backgroundcolor=\color{backcolour},   
    commentstyle=\color{blue},
    keywordstyle=\color{codegreen},
    numberstyle=\tiny\color{codegray},
    stringstyle=\color{codepurple},
    basicstyle=\ttfamily\small,
    breakatwhitespace=false,         
    breaklines=true,                 
    captionpos=b,                    
    keepspaces=true,                 
    numbers=left,                    
    numbersep=5pt,                  
    showspaces=false,                
    showstringspaces=false,
    showtabs=false,                  
    tabsize=2
}
 
\newcommand{\gev}{\; \hbox{GeV}}

\newcommand{\met}{\not\!\!\! E_T}

\def\gev{\, {\rm GeV}}
\def\mev{\, {\rm MeV}}
\def\kev{\, {\rm keV}}

\newcommand{\beq}{\begin{equation}}
\newcommand{\eeq}{\end{equation}}
\newcommand{\be}{\begin{equation}}
\newcommand{\ee}{\end{equation}}
\newcommand{\bea}{\begin{eqnarray}}
\newcommand{\eea}{\end{eqnarray}}

\newcommand{\HyperOpt}{\texttt{Hyperopt }}

\title{ {\color{blue} Collider Detection of Dark Matter Electromagnetic Anapole Moments}}

\author{Alexandre Alves$^a$,}
\author{A. C. O. Santos$^{b,c}$,}
\author{Kuver Sinha$^{d}$}

\affiliation{$^a$Departamento de F\'isica, Universidade Federal de S\~ao Paulo, Diadema-SP, 09972-270, Brazil}
\affiliation{$^b$Departamento de F\'isica, Universidade Federal da Para\'iba, Jo\~ao Pessoa-PB, 58051-970, Brazil \\
$^c$Centre for Cosmology, Particle Physics and Phenomenology (CP3), Université catholique
de Louvain, B-1348, Louvain-la-Neuve, Belgium}
\affiliation{$^d$Department of Physics and Astronomy, University of Oklahoma, Norman, OK 73019, USA}

\emailAdd{aalves@unifesp.br}
\emailAdd{kuver.sinha@ou.edu}

\abstract{Dark matter that interacts with the Standard Model by exchanging photons through higher multipole interactions occurs in a wide range of both strongly as well as weakly coupled hidden sector models. We study the collider detection prospects of these candidates, with a focus on Majorana dark matter that couples through the anapole moment. The study is conducted at the effective field theory level with the mono-$Z$ signature incorporating varying levels of systematic uncertainties at the high-luminosity LHC. The projected collider reach on the anapole moment is then compared to the reach coming from direct detection experiments like LZ. Finally, the analysis is applied to a weakly coupled completion with leptophilic dark matter.}
    
\begin{document} 

\begin{flushright}
OU-HEP-171030
\end{flushright}

\maketitle
\flushbottom
    

\section{Introduction}

One of the defining features of dark matter (DM) models is the nature of the interaction between the DM candidate and Standard Model (SM) particles. This interaction is usually assumed to be mediated by heavy states (for example the SM $Z$ or Higgs boson \cite{Burgess:2000yq, Queiroz:2014yna}, or $Z^{\prime}$ bosons \cite{Alves:2013tqa, Alves:2015dya} belonging to extensions of the SM), although in recent times there has been a surge of interest in hidden-sector models in which the interaction is mediated by new dark forces \cite{Battaglieri:2017aum}.

A class of models with a long history that lies somewhat between these two options is one in which DM interacts with the SM by exchanging photons through higher multipole interactions \cite{Raby:1987ga} - \cite{Barger:2010gv}. At dimension 5 and 6, the effective operators for multipole interactions of a DM fermion $\chi$ can be written as follows
\be \label{EFTops}
\mathscr{L} \supset
    \frac{d_M}{2} \bar\chi \sigma^{\mu\nu} \chi \, F_{\mu\nu}
    + \frac{d_E}{2} \, \bar\chi \sigma^{\mu\nu} \gamma^5 \chi \, F_{\mu\nu} \\
    + \mathcal{A} \, \bar\chi \gamma^\mu \gamma^5 \chi \, \partial^\nu F_{\mu\nu}  \,.
\ee
Here, $d_M$, $d_E$, and $\mathcal{A}$ denote the magnetic, electric, and anapole moment, respectively. For Majorana DM, only the anapole operator is non-zero and can be written in effective field theory (EFT) as
\bea \label{anapoleEFT}
\mathcal{A} &\equiv & \frac{g}{\Lambda^2} \nonumber \\
\mathscr{L}_\text{eff,anapole} &=& \frac{g}{\Lambda^2} \, \bar\chi \gamma^\mu \gamma^5 \chi \, \partial^\nu F_{\mu\nu} \,\,,
\eea
where $\Lambda$ is the cutoff scale. 

These effective descriptions have been studied in a variety of UV completions - for example, models of technicolor \cite{Bagnasco:1993st, Banks:2010eh}, composite DM \cite{Antipin:2015xia},  supersymmetry \cite{Raby:1987ga} and, recently, simplified models of leptophilic DM \cite{Kopp:2014tsa}, \cite{Sandick:2016zut}. In weakly coupled completions, DM is assumed to couple at tree level to heavy charged states and hence at one loop to the photon. Restricted to the case of supersymmetry, this could be a model of Bino DM coupling to sleptons \cite{Dutta:2014jda}. In strongly coupled completions, the DM candidate can be a composite state of charged particles. The cutoff scale $\Lambda$ corresponds to either the mass of heavy charged states running in the loop, or the scale of confinement in the strongly coupled hidden sector.

The purpose of this paper is to explore the detection prospects at the HL-LHC of DM with couplings to the SM shown in Eq.~(\ref{EFTops}). We focus, in particular, on the anapole operator of Eq.~(\ref{anapoleEFT}) and calculate the reach of the high luminosity LHC (HL-LHC) in probing the cutoff scale $\Lambda$ for different DM masses. A collider study of electric and magnetic dipole moments is left for future work\footnote{We note that collider studies of magnetic dipole DM have been performed in \cite{Primulando:2015lfa} and \cite{Kadota:2014mea}, for both the LHC and the ILC, using techniques different from the ones employed in our paper.}. Our collider study is conducted in the conservative and comparatively clean mono-$Z$ channel, incorporating varying levels of systematic uncertainties. We utilize analysis methods developed recently by a subset of the authors. After carefully choosing kinematic variables that can discriminate between signal and SM background in Section \ref{ColliderStudy}, we select cuts using the Bayesian optimization method implemented in the Python algorithm \HyperOpt~\cite{hyperopt}. A Boosted Decision Tree (BDT) is then used to classify events into signal and background classes, after a joint optimization of kinematic cuts and BDT hyperparameters.

We have two main motivations for this study. The first is to connect to the rather extensive body of literature on direct detection prospects for this class of DM models. Electromagnetic anapole and dipole DM has been investigated in the context of experiments like DAMA, CDMS, XENON and LUX by several groups \cite{DelNobile:2014eta} - \cite{Gresham:2013mua}, and projections for the future LZ experiment based on a simple scaling of the scattering cross section has been given in \cite{Kopp:2014tsa}, \cite{Sandick:2016zut}. At the level of the EFT, the reach in the cutoff scale $\Lambda$ obtained from our collider study can be mapped on to a reach in the value of the anapole moment $\mathcal{A}$ through Eq.~(\ref{anapoleEFT}) (taking dimensionless couplings $g \sim \mathcal{O}(1)$). Since the anapole moment determines the scattering cross section of DM off nuclei, the collider reach can then be compared to the reach of DM direct detection experiments. We do this in Fig.~\ref{dmcolliderresults}, comparing the values of the anapole moment probed by LUX and LZ to the values probed by our collider study, assuming 5, 10, and 20 $\%$ systematic uncertainties at the HL-LHC with 3000 fb$^{-1}$ of data.

A second motivation for our paper lies in applying this comparative study to a particular UV completion. Over the last few years, there has been steady interest in models of leptophilic DM interacting with the SM via heavy charged mediators \cite{Fukushima:2014yia} - \cite{Garny:2015wea}. A comprehensive one-loop analysis of the direct detection phenomenology of this class of models has been performed by \cite{Ibarra:2015fqa}. The relic density and indirect detection rates have been calculated by \cite{Ibarra:2015fqa} and their embedding within supersymmetry has been studied by \cite{Fukushima:2014yia}. In \cite{Sandick:2016zeg}, constraints on this class of models were obtained under the assumption of a DM spike near the supermassive black hole at the center of our Galaxy. For small mass separation between the charged mediators and the DM candidate, these models are difficult to probe at the LHC by direct production of the mediators themselves \cite{Liew:2016oon}, \cite{Delannoy:2013ata}. On the other hand, an explicit calculation reveals that the anapole moment is enhanced precisely in these compressed regions of parameter space. Since the coupling to the photon increases, we obtain a corresponding enhancement in the performance of our collider study, as well as the scattering cross section with nuclei. The expectation, then, is that our collider study, in conjunction with projections from LZ, should be able to probe these compressed parameter regions. This aspect of our study is conducted in Section \ref{appsimpmodel}.

The paper is structured as follows. After performing our collider analysis in Section \ref{ColliderStudy} - Section \ref{BDTclassif}, the results of our EFT analysis are displayed in Fig.~\ref{reachLambda} in terms of the cutoff scale $\Lambda$, and in Fig.~\ref{dmcolliderresults} in terms of the anapole moment $\mathcal{A}$. In the latter figure, the limits from current and future direct detection experiments are also displayed, following a discussion of the methods used to calculate those limits in Section \ref{directdetc}. The EFT results are then applied to the case of a simplified model with charged mediators in Section \ref{appsimpmodel}. We end with our Conclusions.

\section{Collider Study} \label{ColliderStudy}

In this section, we first provide a brief overview of prior work on multipole DM. We then present the results of our collider study.

The relic density of anapole and dipole DM has been worked out by many authors \cite{Ho:2012bg} - \cite{Ibarra:2016dlb}. In particular, \cite{Gao:2013vfa} calculated the relic density in the DM mass range $m_{\chi} \sim \mathcal{O}(100-500)$ GeV, incorporating annihilation channels like $\chi \chi \rightarrow W^+ W^-$ and $\chi \chi \rightarrow t \overline{t}$. For DM with $m_{\chi} \sim 100$ GeV, the correct thermal relic density is obtained for a value of the cutoff scale $\Lambda \sim 700$ GeV. While noting, from our Fig.~\ref{reachLambda}, that this critical value of $\Lambda$ will be probed at the HL-LHC, we will in general remain agnostic about the relic density constraint. Depending on the cosmological history of the Universe prior to Big Bang Nucleosynthesis, a wide range of annihilation cross sections can in any case be allowed \cite{Kane:2015jia}, \cite{Dutta:2009uf}. 

Low mass multipole DM has been studied by several groups in the context of anomalies in direct detection experiments \cite{DelNobile:2014eta} - \cite{Gresham:2013mua}. We note that low mass multipole DM has also found applications in addressing the longstanding Solar Abundance Problem (discrepanies between solar spectroscopy and helioseismology) \cite{Geytenbeek:2016nfg}. The preferred anapole moment in such models turns out to be $\Lambda \sim \mathcal{O}(1)$ GeV$^{-2}$. However, in our current paper, we restrict our attention to DM with mass $\sim \mathcal{O}(100)$ GeV. Collider searches for low mass anapole or dipole DM will require different techniques that are left for future investigation. 

We now turn to a discussion of the collider prospects of anapole DM with an emphasis on mono-$X$ channels, where $X = j, \, \gamma, \, Z, \, h$ \cite{Goodman:2010ku} - \cite{Goodman:2010yf}. These channels can serve as fertile places to search for anapole DM at colliders. The only requirement is that initial or final states contain charged particles which can emit a photon, which will then ultimately couple to the anapole DM. To our knowledge, the earliest appraisal of anapole DM in the context of the LHC appeared in \cite{Gao:2013vfa}. The authors performed a monojet study with 19.5 (10.5) fb$^{-1}$ of  CMS (ATLAS) data at 8 TeV and obtained bounds on the cutoff scale $\Lambda \sim 350$ GeV. We note that the monojet cross section is expected to dominate over the mono-$Z$ signature for this class of models. Moreover, the mono-$Z$ channel suffers from the usual branching ratio penalty of demanding $Z$ decay to leptons. On the other hand, though, we also expect that systematic uncertainties on the background should scale more favorably for the mono-$Z$ process at the HL-LHC. A comparative study of mono-$Z$ and monojet signatures for this class of models at the HL-LHC is left for future work. For now, our collider results based on the mono-$Z$ signature should be treated as a conservative estimate.

The mono-$Z$ channel, moreover, offers a good compromise between the signal production cross section on the one hand, and the information available in the $Z$ boson decay on the other. Our strategy will be to use the final state particle distributions to train a decision tree algorithm in order to efficiently separate signal and background events. While allowing $Z$ boson decays to jets would increase the number of signal events, the SM background associated with jets plus missing $E_T$ is large.  On the other hand, the leptonic decay mode is a viable alternative since it is a cleaner channel and the associated cross section is not too much smaller. Previous searches for dark matter in the mono-$Z$ channel with and without machine learning tools showed good discovery prospects~\cite{Alves:2015dya, Carpenter:2012rg}.

We therefore perform a mono-$Z$ search in the leptonic channel at the 13 TeV LHC. Our signal is
\begin{equation}
pp\to Z+\gamma^*\to \ell^+\ell^- + \chi\overline{\chi}
\label{mainprocess}
\end{equation}
where $\ell=\mu,e$ come from the $Z$ boson and the dark matter pair from the virtual photon. The backgrounds considered in this work are the main irreducible ones
\begin{itemize}
\item  $ZZ(\gamma^*) \to \ell^+\ell^- + \nu_\ell\overline{\nu}_\ell$, and 
\item  $W^+W^- \to \ell^+\ell^{\prime -} + \nu_\ell\overline{\nu}_{\ell^\prime}$, 
\end{itemize}
and the main reducible ones 
\begin{itemize}
\item $ZW \to \ell^\pm\ell^\mp\ell^{\prime\pm} + {\nu}_{\ell^\prime}$ and 
\item $t\bar{t} \to W^+W^- b\bar{b}\to \ell^+\ell^{\prime -} + \nu_\ell\overline{\nu}_{\ell^\prime}+jj$. 
\end{itemize}

The irreducible $\tau^+\tau^-$ background is very small after $\tau$ decays to lighter leptons. The single top background $Wt$ has a final state similar to the $t\bar{t}$ background when the $W$ boson and the top quark decay leptonicaly but with a somewhat smaller jet multiplicity. Yet, just like $t\bar{t}$, as we are going to see in the next section, the larger jet multiplicity makes the $Wt$ identification by the BDT classifier very efficient. Because the $Wt$ rate is an order of magnitude smaller that $t\bar{t}$, it can be safely neglected.

We require the following basic selection criteria for the mono-$Z$ events:
\begin{equation}
p_T(\ell) > 20\; \hbox{GeV},\;\; |\eta_\ell| < 2.5,\;\; \Delta R_{\ell\ell}>0.4,\;\; \met > 20\; \hbox{GeV} 
\label{eq:cuts}
\end{equation}
two opposite-charge leptons (electrons or muons) with transverse momentum greater than 20 GeV in the central region of the calorimeter and missing energy larger than 20 GeV for trigger purposes. These initial cuts are loose since ultimately we will tune the $\met$ cut concurrently with tuning the hyperparameters of the machine learning (ML) algorithm. This approach proved to be very efficient in optimizing the performance of the decision trees algorithm in SM double Higgs production~\cite{Alves:2017ued}, a project undertaken recently by some of the authors. 

We also do not demand an explicit jet veto at this point. Instead, we pass the number of reconstructed jets and leptons to the decision trees algorithm in order to facilitate the identification of the reducible backgrounds, as discussed in the next section. The lack of a lepton invariant mass cut may be noticeable as well. The task of selecting the events by cutting on kinematic variables is the job of the BDT and may be delegated entirely to the ML training phase. We actually found that the BDT performs better when we keep the pre-selection of events at a minimum.

The DM effective operator in Eq.~(\ref{anapoleEFT}) was implemented in \texttt{FeynRules}~\cite{Alloul:2013bka}. Events were generated with \texttt{MadGraph5}~\cite{Alwall:2011uj} at leading order with one extra jet. The hard and soft jet regimes were matched in the MLM scheme~\cite{mlm} at appropriate separation scales. Hadronization was performed with \texttt{Pythia6}~\cite{pythia}. For the detector simulation and jet clustering we used \texttt{Delphes3.3}~\cite{delphes} and \texttt{FastJet}~\cite{Cacciari:2011ma} with the anti-$kt$ algorithm. The luminosity was fixed at 3 ab$^{-1}$ projecting the results to the end of the experiment.

The matched cross sections of signal and background processes after the basic selection criteria are displayed in table~\ref{tab:1} for DM masses from 100 to 500 GeV and $\Lambda=1$ TeV at the 13 TeV LHC.

\begin{table} 
\label{tab:1}
\centering
\begin{tabular}{c|c|c|c|c|c}
\hline
\hline
Signals & 100 GeV & 200 GeV & 300 GeV & 400 GeV & 500 GeV \\
\hline
$\sigma$(fb) & 0.143 & 0.119 & 0.095 & 0.073 & 0.056 \\
\hline
\hline
Backgrounds & $ZZ$ & $WW$ & $ZW$ & $t\bar{t}$ & $Wt$ \\
\hline
$\sigma$(fb) & 152.4 & $1.5\times 10^3$ & 236.2 & $1.4\times 10^4$ & 584.9\\
\hline 
\hline
\end{tabular}
\caption{Signal and main background cross sections after basic cuts of Eq.~(\ref{eq:cuts}) in fb at the 13 TeV LHC. The $Wt$ background turns out to be negligible after the BDT classification.} 
\end{table}

\section{Kinematic Variables for BDT Discrimination} \label{variables}

In this Section, we describe the kinematic variables used to represent our simulated data. Each event is represented by a real-valued vector composed of the following ten kinematic variables, inspired by the mono-$Z$ study performed in ref.~\cite{Alves:2015dya}:
\begin{itemize}

\item Missing energy $\met$. This variable is used both for cutting and BDT training. Events with heavier dark matter are characterized by harder $\met$ spectrum.

\item The product $\met \, \times \, \cos\left(\Delta\phi(\vec{E}_T^{miss},\vec{p}_T^Z)\right)$, where $\Delta \phi$ is the angle between the two dimensional vector $\vec{E}_T^{miss}$ and the transverse momentum $\vec{p}_T^Z$ of the $Z$ boson candidate. This variable is a measure of axial-$\met$, which is the projection of $\vec{E}_T^{miss}$ in the direction opposite to the $Z$ candidate~\cite{Aad:2012awa}. It is useful to differentiate among various DM operators in EFT ~\cite{Carpenter:2012rg}.

\item The variable $|\met - p_T^Z|/p_T^Z$, which is the fractional $p_T$ difference~\cite{Aad:2012awa}.

\item The variable $\Delta\phi(\ell^+,\ell^-)$, which is the azimuthal separation of the two leptons.

\item The variable $\alpha_T \, = \, E_T(\ell_2)/M_T$, where $E_{T2}$ is the transverse energy of the softest lepton of the $\ell^+\ell^-$ pair and $M_T=\sqrt{(E_{T1}+E_{T2})^2-(p_{x1}+p_{x2})^2-(p_{y1}+p_{y2})^2}$ ~\cite{Edelhauser:2015ksa}.

\item The variable $\cos(\theta^{*})$~\cite{Barr:2005dz} where $\theta^*$ is defined as the boost invariant $\cos(\theta^*)=\tanh\left(\frac{\eta_{\ell^+}-\eta_{\ell^-}}{2}\right)$. This variable has been used in supersymmetric studies, and is correlated to the production angle of sparticles. It has been studied in decay processes such as slepton to leptons + $\met$~\cite{Barr:2005dz}, or sbottoms to bottom jets + $\met$~\cite{Alves:2007xt}.

\item The variable $MT_c=\sqrt{2\left(\vec{p}_{T_\ell}\cdot\vec{p}_{T_\ell}+p_{T_\ell}p_{T_\ell}\right)}$, which is the contransverse mass \cite{Tovey:2008ui}.

\item The variable $M_{\ell\ell}$, which is the invariant mass of the leptons pairs. It is useful in distinguishing between leptons from $Z$ and $W$ decays.

\item The variable $n_\ell$, which is the number of leptons identified in the event. The majority of $WZ$ events present three charged leptons.

\item The variable $n_j$, which is the number of jets. Leptonic top quark pair production events present at least two hard jets.

\end{itemize}

\begin{figure}[t]
  \centering
  {\includegraphics[width=.48\textwidth]{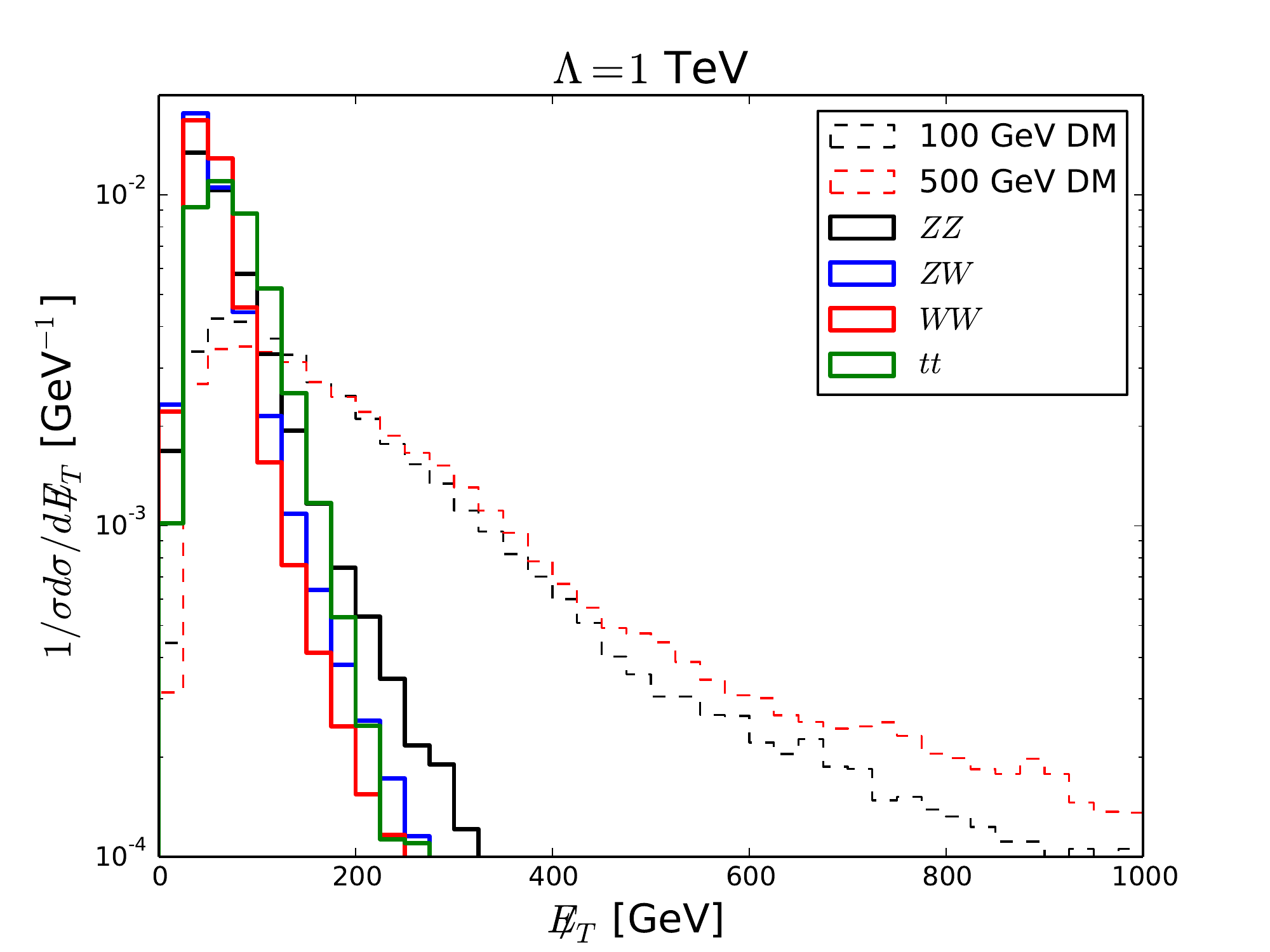}}\quad
  {\includegraphics[width=.48\textwidth]{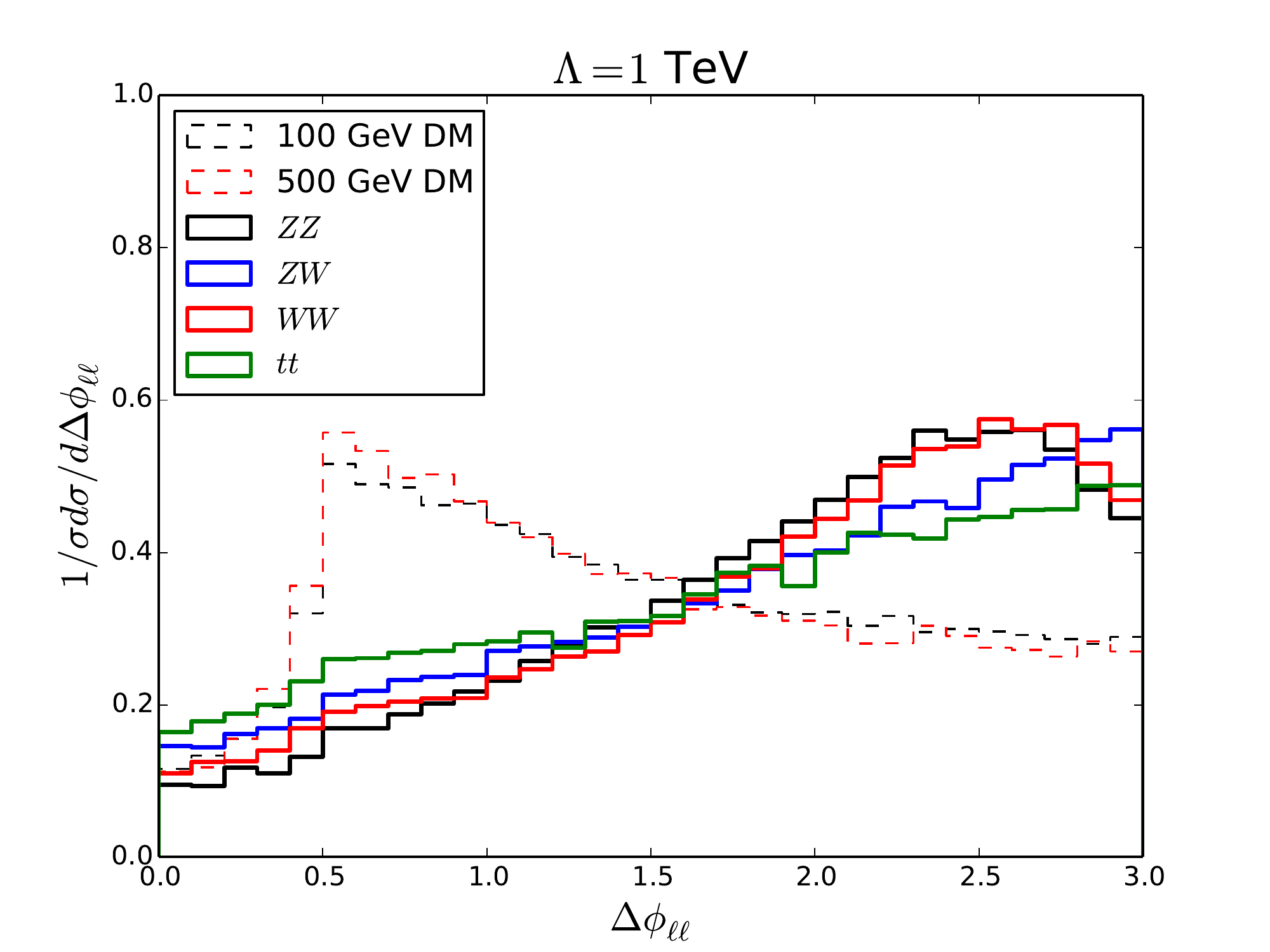}}\\ 
  {\includegraphics[width=.48\textwidth]{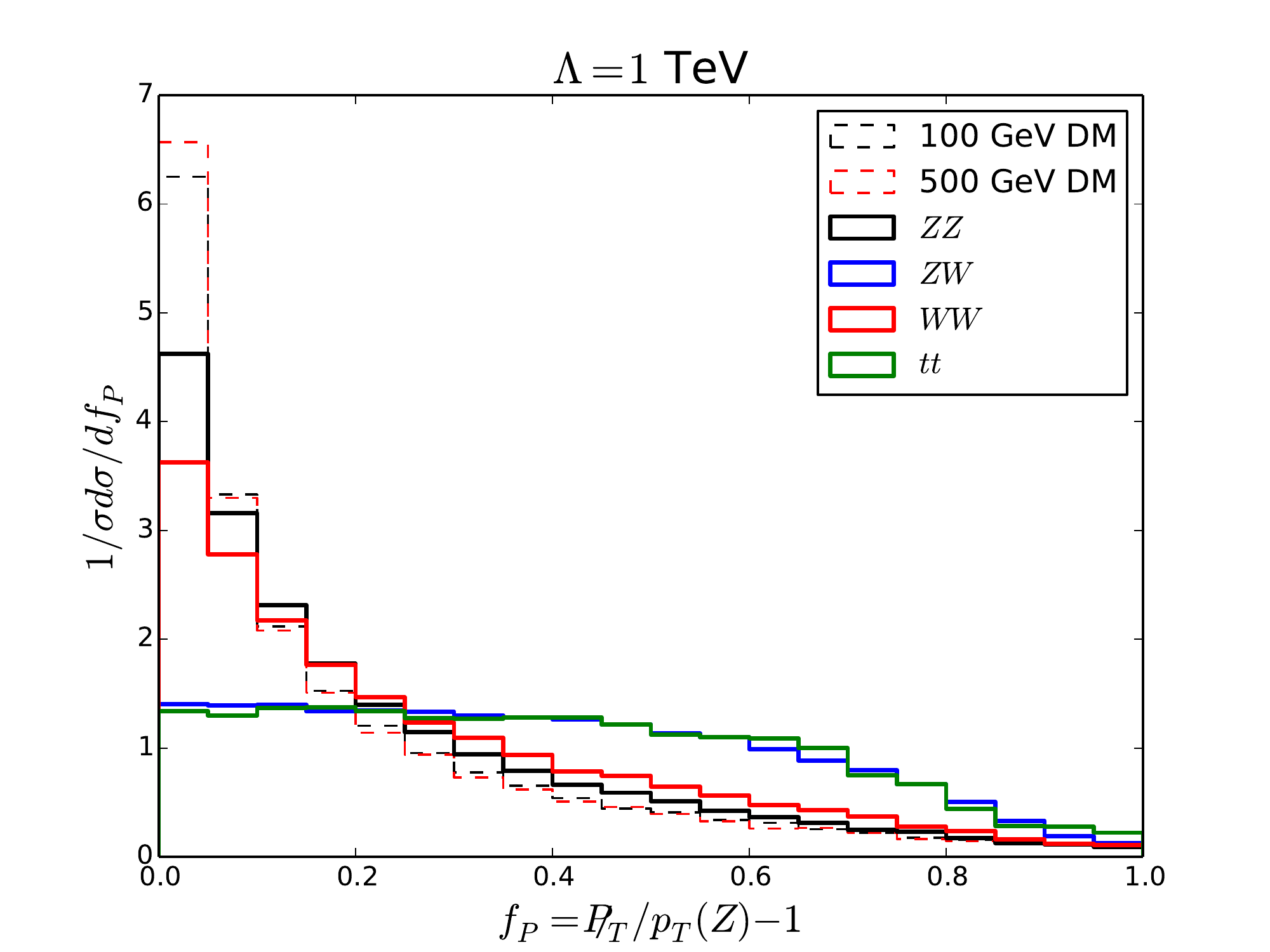}} \quad
  {\includegraphics[width=.48\textwidth]{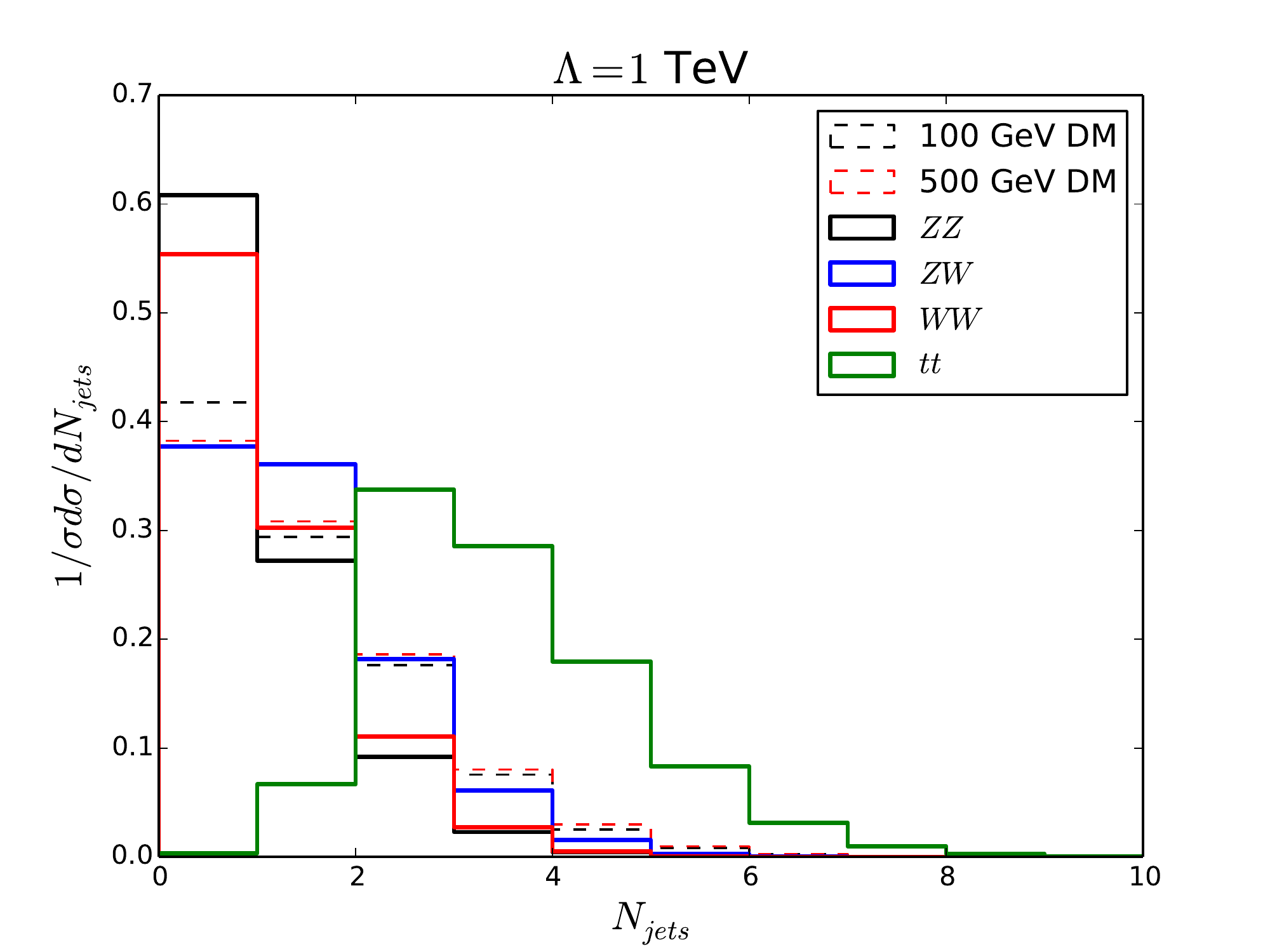}}\\
  \caption{Kinematic variables used in the BDT study: Upper panel: Plot of missing energy $ \,\, \met$ (left) and azimuthal separation of two leptons $ \Delta \phi_{ll}$ (right) distributions. Lower panel: Plot of the fractional $p_T$ difference (left), and the number of jets $ \,\, N_{jets}$ (right). The cutoff scale is fixed at $\Lambda = 1$ TeV. The solid lines correspond to the backgrounds and black (red) dashed lines correspond to DM distributions.
  }  
  \label{fig:1}
\end{figure}

Some of the variables defined above are shown in Fig.~\ref{fig:1}. In the left upper plot we show the missing $E_T$ distribution for the backgrounds (solid lines) and for the lighter (100 GeV) and the heavier (500 GeV) dark matter considered in this work. As expected, the heavier the DM particle is, the harder is its spectrum compared to the softer backgrounds. In the right upper plot, we display the difference of the azimuthal angle of the two hardest leptons in the event. As a consequence of the harder missing transverse momentum, DM events present more collimated leptons from the $Z$ boson decay compared to the backgrounds. Two good variables to discern the reducible backgrounds $ZW$ and $t\bar{t}$ are shown in the lower plots. On the left, we can see that the typical transverse momentum of the lepton pair is balanced by $\not\!\! p_T$ in events containing $Z$ bosons, but less balanced for events containing $W$s. Finally, on the right is displayed the number of hard jets identified in the event. Events with top quarks contain at least two hard jets and a considerably large fraction of events with higher jet multiplicity compared to the other processes. As observed in ref.~\cite{Alves:2015dya}, the other variables not shown in Fig.~\ref{fig:1} present good discrimination power as well, especially the lepton invariant mass. In contrast to ref.~\cite{Alves:2015dya}, however, we use all these variables to represent our simulated data in the learning process of the BDT classifier.

\section{BDT Classification and Performance} \label{BDTclassif}

In this Section, we discuss our decision tree analysis and give our results.

We utilized the  \texttt{XGBoost} package~\cite{xgboost} to train boosted decision trees. Approximately 1.5 million events were generated, with around 300,000 for each event class - i.e., the signal class and the four background classes. One hundred thousand events were singled out for optimization purposes and the remaining to train/validation and test the BDT in the proportion of 2/3 and 1/3 of the events, respectively.

Since the DM mass is not known at the stage when the mono-$Z$ signal is studied, it is not possible, in principle, to optimize the ML algorithm to discern the signal for a given mass. Yet, the DM mass in the signal class needs to be fixed to train the BDT. We chose to fix the DM mass at 100 GeV, the initial value of our mass scan. This choice was motivated by the fact that a 100 GeV particle signal is harder to discern from the background than a heavier one, as can be observed from the distributions shown in Fig.~\ref{fig:1}. Our expectation is that the algorithm also present a good performance for the heavier (and easier) dark matter signals. We checked that, in fact, heavier dark matter is more  easily identified  as a signal event in this approach. A more sophisticated approach, based on parameterized neural networks, is also possible~\cite{Baldi:2016fzo}.

\begin{figure}[h]
  \centering
  \includegraphics[width=.8\textwidth]{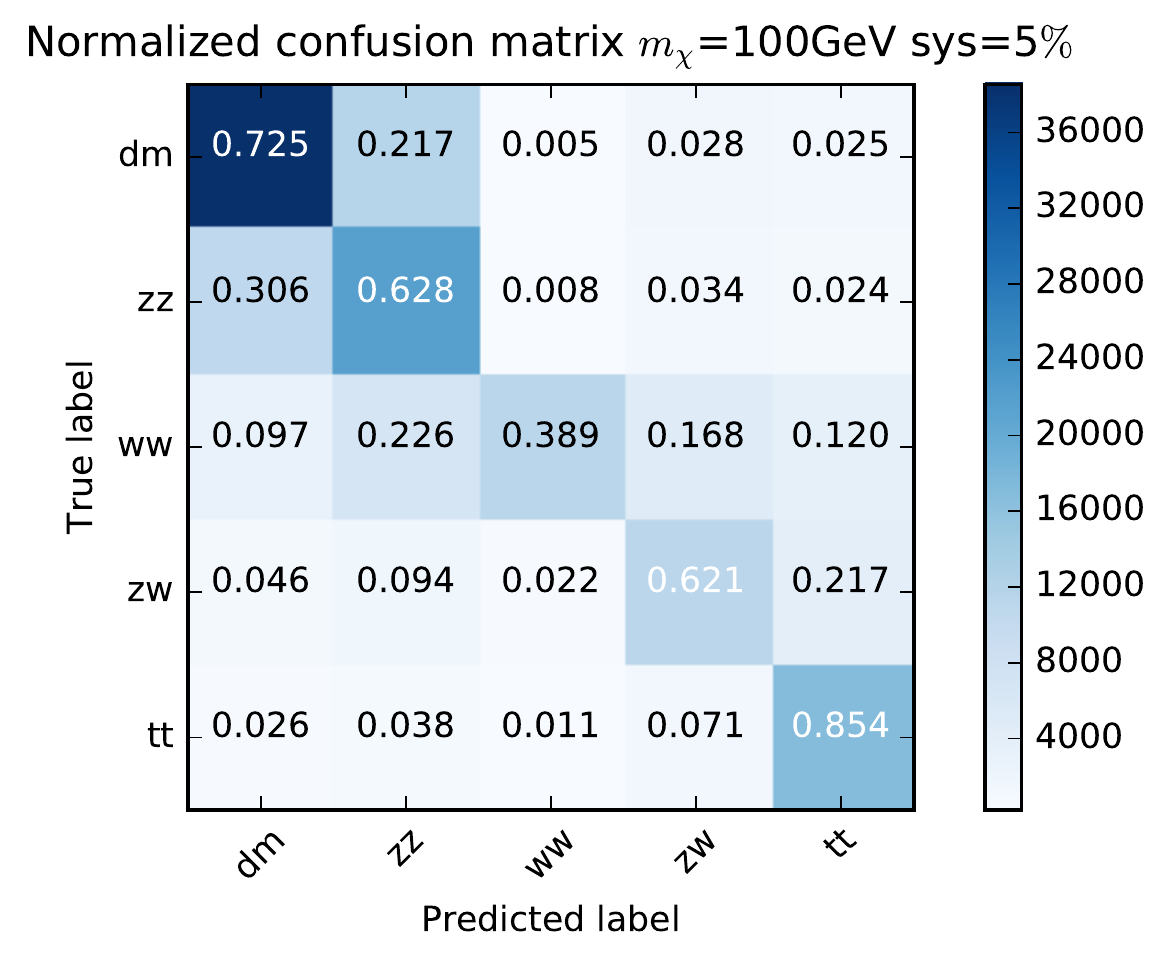}
  \caption{BDT performance plot: Normalized confusion matrix for the multi-channel analysis at $m_{\chi} = 100$ GeV, and systematic uncertainties of $5\%$ for the test samples. The grid consists of the signal class (labeled ``dm"), and the four major background classes (labeled $ZZ$, $ZW$, $WW$, and $tt$) corresponding to the backgrounds discussed below Eq.~(\ref{mainprocess}). It is evident that around 22\% of the signal events are predicted to be $ZZ$ events and 30\% of $ZZ$ events are predicted to be signal events.}
\label{confusion}
\end{figure}

The BDT hyperparameters, the $\met$ threshold cut, and the number of leptons and jets in order a given event be vetoed were all adjusted jointly in a Bayesian optimization framework with \texttt{HyperOpt}~\cite{hyperopt}. The joint optimization of cuts and ML hyperparameters is advantageous once the kinematic cuts affect the performance of these algorithms in way that is hard to predict. Reducing the number of background events with hard cuts help to increase the signal significance, but has a deleterious effect on the BDT classification as the kinematic distributions of the various classes become more similar to each other. Delegating all the discrimination to the ML side, on the hand, might not suffice if the backgrounds are too much larger than the signal. The optimum point in this trade off is achieved by the joint optimization as described in details in ref.~\cite{Alves:2017ued}.

 The signal significance was calculated taking a systematic uncertainty in the total backgrounds yield, $\varepsilon_B$. We estimated four scenarios: one with almost no systematic uncertainties, taking $\varepsilon_B=$1\%, and several others with varying degrees of uncertainties, taking $\varepsilon_B=$ 5, 10 and 20\%. The joint optimization described in the previous paragraph was performed taking these systematic uncertainties into account, in order that the optimization algorithm learns to increase the $S/B$ ratio and tame the effects of these uncertainties as explained in detail in ref.~\cite{Alves:2017ued}.

\begin{figure}[h]
  \centering
    \includegraphics[width=.8\textwidth]{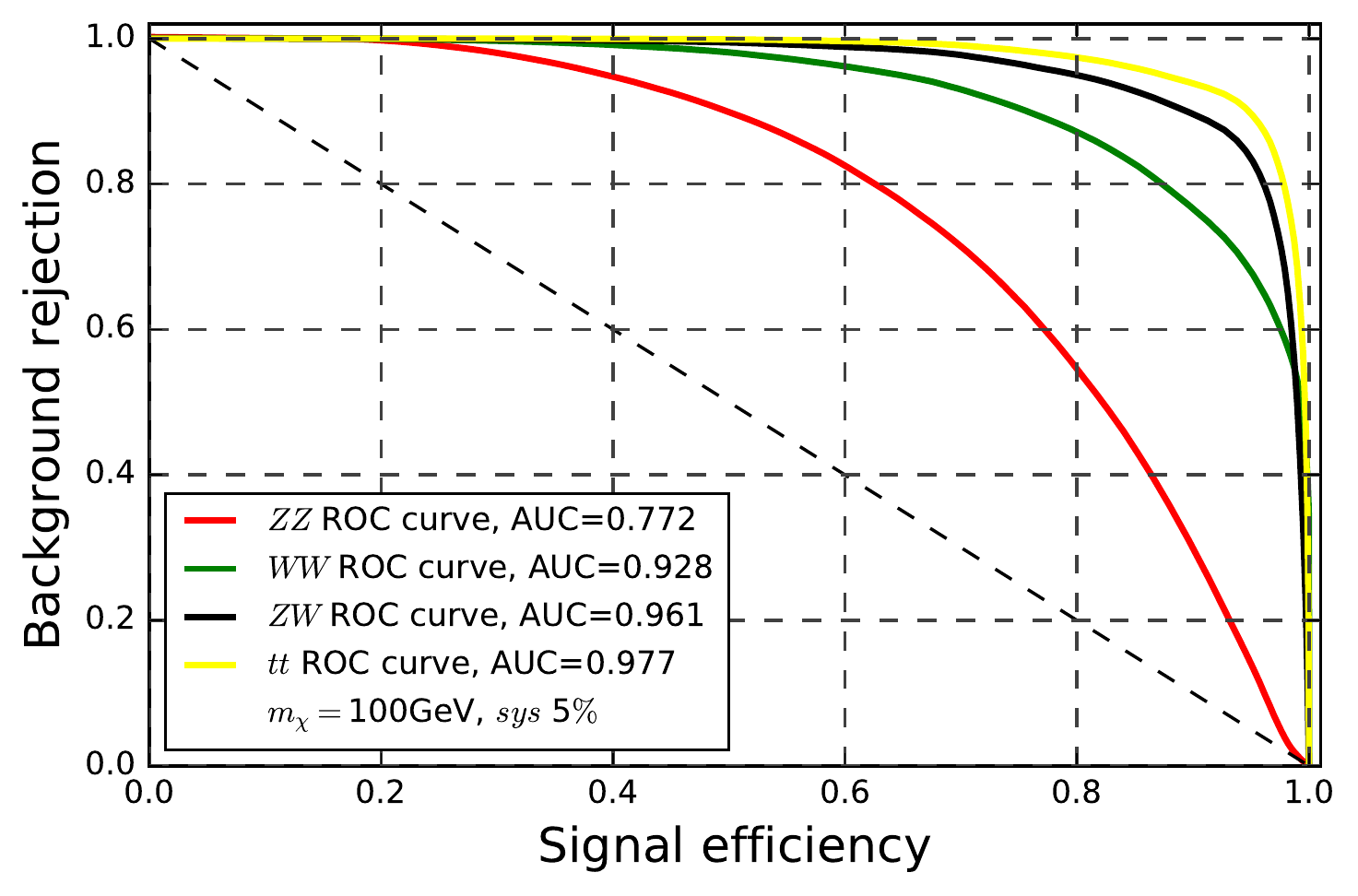}
     \caption{BDT performance plot: Receiver Operating Characteristic (ROC) curves for the multi-channel analysis at $m_{\chi} = 100$ GeV, and systematic uncertainties of $5\%$ for the test samples. The red, green, blue, and yellow curves show the $ZZ$, $WW$, $ZW$, and $t \bar{t}$ backgrounds, respectively. The $ZZ$ background shows the lowest area under curve (AUC). The $t\bar{t}$ has a larger background rejection rate due its larger jet multiplicity.}
     \label{rocplot}
\end{figure}  

We now show the performance plots for our BDT analysis. We will discuss, in turn, the confusion matrix, the Receiver Operating Characteristic (ROC) curves, the BDT output distributions for signal and backgrounds, and, finally, the $\Lambda$ scale that can be probed with 3 ab$^{-1}$ of data at the 13 TeV LHC, all obtained for the test samples.
  
In Fig.~\ref{confusion}, we show the normalized confusion matrix for the multi-channel analysis at $m_{\chi} = 100$ GeV, and systematic uncertainties of $5\%$. The grid consists of the signal class (labeled ``dm"), and the four major background classes (labeled $ZZ$, $ZW$, $WW$, and $tt$) corresponding to the backgrounds discussed below Eq.~(\ref{mainprocess}). As expected, it is the $ZZ$ background that is most difficult to discern from the signal. From the confusion matrix we see that around 22\% of the signal events are predicted to be $ZZ$ events and 30\% of $ZZ$ events are predicted to be signal events. The reducible $ZW$ and $t\bar{t}$ are easily identified as background events due their larger lepton and jet multiplicities, respectively, as anticipated in the previous section.

In Fig.~\ref{rocplot}, we show the  Receiver Operating Characteristic (ROC) curves for the multi-channel analysis at $m_{\chi} = 100$ GeV, and systematic uncertainties of $5\%$. The red, green, blue, and yellow curves show the $ZZ$, $WW$, $ZW$, and $t \bar{t}$ backgrounds, respectively. The results of the confusion matrix are corroborated by the ROC curves of Fig.~\ref{rocplot}. Indeed, it is the $ZZ$ background that shows the smallest area under curve (AUC), signifying the smallest background rejection for a given signal acceptance. We note, on the other hand, that $t\bar{t}$ has a larger background rejection rate, mainly because of its larger jet multiplicity. The lepton multiplicity helps to discern the $ZW$ backgrounds which has the second largest AUC. We found that the optimum $\met$ cut varies between 50 to 90 GeV depending on the systematics level and no jet or lepton vetoes. That is, the best performance was achieved by delegating the task of enhancing the classification performance more to the BDT, and less to the kinematical cuts.
%
%
 
\begin{figure}[t]
  \centering
    \includegraphics[width=.8\textwidth]{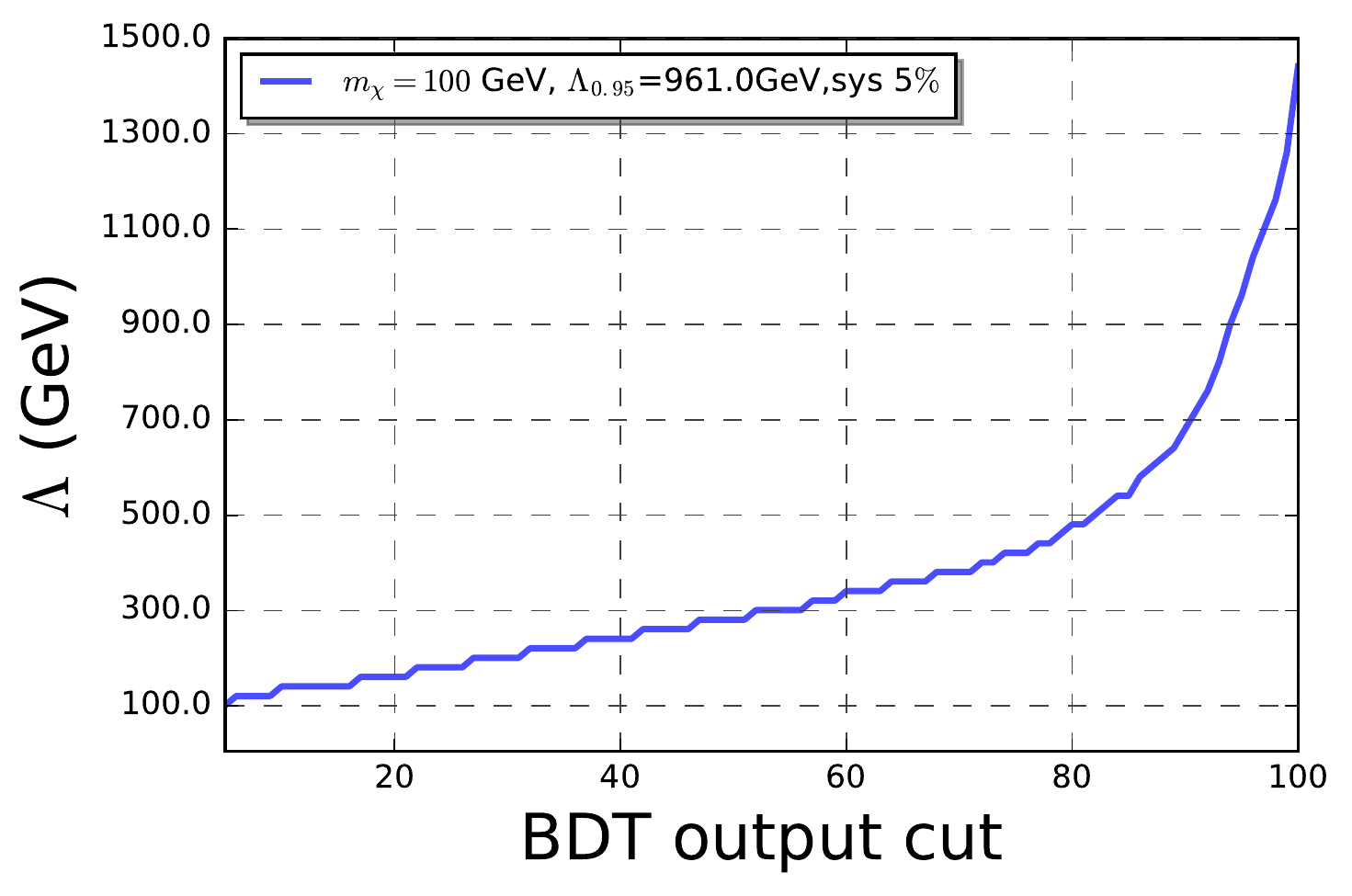}
  \caption{BDT performance plot: The cutoff energy scale $\Lambda$ for discovery (5$\sigma$) as a function of BDT score cut for a 100 GeV dark matter and assuming a 5\% systematics. We chose a cut of 0.95 in the BDT output to separate signal and backgrounds.}  
  \label{lambdaversusbdt}
\end{figure}

In Fig.~\ref{lambdaversusbdt}, we plot the cutoff energy scale $\Lambda$ for discovery (5$\sigma$) as a function of BDT score cut. We see that the $\Lambda$ scale which can be probed by the 13 TeV LHC increases very rapidly as the output cut approaches 1. We chose to keep the cut at 0.95 for the sake of the stability of the results. A harder cut probes regions with too few background events, which leads to larger fluctuations. In order to estimate the reach in $\Lambda$, we averaged the results of ten runs, randomly reshuffling the train and test samples at each run. The uncertainty of the LHC reach in $\Lambda$ is around $\pm$10-15\% of the estimated $\Lambda$ depending on the mass at the 0.95 output threshold. The results shown in Fig.~\ref{lambdaversusbdt} are from this averaging process, keeping a 5\% systematics level. Results for other DM masses and systematic errors were obtained in the same way. After the BDT output cut, 31(18) signal and 136(164) background events survive for a 100(500) GeV dark matter. 

\begin{figure}[!t]
  \centering
  \includegraphics[width=.8\textwidth]{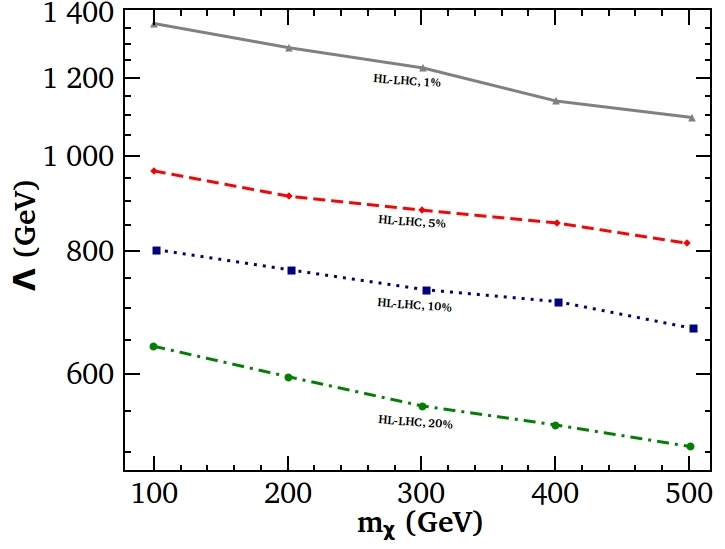}
  \caption{Collider results: The $5 \sigma$ reach in the cutoff energy scale $\Lambda$ as a function of dark matter mass  for several values of systematics uncertainties. The luminosity is fixed at 3000fb$^{-1}$.}  
  \label{reachLambda}
\end{figure}

In Fig.~\ref{reachLambda}, we display the $5 \sigma$ reach in the cutoff energy scale $\Lambda$ as a function of DM mass for several values of systematics uncertainties. The luminosity is fixed at 3000fb$^{-1}$. The performance of the BDT classifier improves as the dark matter gets heavier. This behavior can be understood when we look at the distributions of Fig.~\ref{fig:1}. For example, the 500 GeV DM presents kinematic distributions which are less similar to the backgrounds. Most importantly, the signal distribution in that case is distinct from the $ZZ$ background, making the BDT classification more efficient. Of course, as the cross section decreases with the DM mass, the $\Lambda$ scale that can be actually probed drops with the DM mass as shown in Fig.~\ref{reachLambda}. In the case of the 300 GeV mass, the reach of the scale $\Lambda$ only changed slightly compared to the 200 GeV case - a persistent effect up to the 10\% systematics level. In this case, the gain in the BDT classification is competitive with the drop in cross section. For larger masses, however, the number of signal events produced was not enough to beat the better classification achieved with the ML algorithm and the estimated LHC sensitivity in $\Lambda$ is degraded. 


Assuming a very small systematic uncertainty of 1\%, the scales for which the anapole DM can be discovered, at $5\sigma$, are all above $\Lambda \sim 1.1$ TeV, as can be seen in Fig.~(\ref{reachLambda}). Assuming systematic uncertainties at the level of 5\%, a 100 GeV DM can still be discovered if $\Lambda \approx $ 1 TeV. From 5\% to 10\% and from 10\% to 20\% systematics, the discovery reach in $\Lambda$ decreases approximately by 200 GeV for a given DM mass. 

\section{Comparison with Direct Detection Limits} \label{directdetc}

In this Section, we compare the collider constraints on the anapole moment to constraints coming from direct detection experiments. 

We are interested in DM masses $\mathcal{O}(100\gev)$ and typical nuclear recoil energy $\sim10$-$30\kev$. This corresponds to a DM-nucleus momentum transfer
\begin{equation*}
\sqrt{|q^{2}|}\approx|\pmb{q}|\approx 50\sim 80\mev,
\end{equation*}
where $\pmb{q}$ is the three-vector of the four-momentum $q$. We assume that the DM has a coupling with the electromagnetic field given by Eq.~(\ref{anapoleEFT}). At small momentum transfer, the interaction of the DM and the nucleus can be described by the following effective Lagrangian
\begin{equation}
\label{eq:analag}
\mathcal{L}_{\text{DM-nucleus}}=\frac{i\mathcal{A}}{2}\overline{\chi}\gamma^{\mu}\gamma^{5}\chi\partial^{\nu}F_{\mu\nu}+eA_{\mu}J^{\mu},
\end{equation}
where  $J^{\mu}$ is the nuclear current operator. We are neglecting the  $q^{2}$ dependence in $\mathcal{A}$; a detailed calculation performed in a previous paper by a subset of the authors revealed that this introduces at most  a $0.6\%$ error in the anapole moment for the energies considered here \cite{Sandick:2016zut}.

The differential cross section for the scattering of DM with nuclei is given by ~\cite{Kopp:2014tsa, DelNobile:2014eta, Ho:2012br, Gresham:2013mua}
\be
\frac{d\sigma}{dE_{R}} = 4 \alpha_{\text{em}}\mathcal{A}^{2}Z^{2}F^{2}_{Z}(\pmb{q}^{2})\left[2m_{T}-\left(1+\frac{m_{T}}{m_{\chi}}\right)^{2}\frac{E_{R}}{v^{2}}\right] +  4 \mathcal{A}^2 d_A^2 F_s^2 \bigg(\frac{J+1}{3J}\bigg)
              \frac{2 E_R m_T^2}{\pi v^2} \,\,.
              \label{eq:dsdE-anapole}
\ee
Here, the mass of the target nucleus is denoted by $m_{T}$, $E_{R}=\pmb{q}^{2}/(2m_T)$ denotes the recoil energy of the nucleus, $Z$ the nuclear charge, and $v$ the velocity of the DM particle. $F_{Z}$ is the nuclear charge form factor, while the nuclear spin form factor is denoted by $F_s$. The second term corresponds to scattering off the nuclear dipole moment $d_A$, which is small for Xenon. 

The differential rate per unit target mass is 
\bea
  \frac{dR}{dE_R} &= \frac{\rho_0}{m_\chi m_T}
    \int_{v_\text{min}}^\infty \! d^3v \frac{d\sigma}{dE_R} v \, f_\oplus(\vec{v}) \,\,,
\eea
where, $\rho_0$ is the local DM density. The minimal velocity of DM that is required for a recoil energy $E_R$ is given by 
$v_{\hbox{min}} = \sqrt{m_T E_R /2} / M_{\rm red}$, where $M_{\rm red}$ is the reduced mass of the DM-nucleon system. The DM velocity distribution in the rest frame of the detector is given by  $f_\oplus(\vec{v})$.

Based on these calculations, several groups have calculated the scattering cross section of multipole DM. The constraints can be depicted as upper bounds on the anapole or dipole moments, and we will mainly use the results obtained in the $m_{\chi} \sim \mathcal{O}(100)$ GeV range. The constraint on the scattering cross section for a given DM mass obtained from LUX 2016 \cite{Akerib:2016vxi} can be scaled to the corresponding projected constraint at LZ. We will take the most optimistic scenario, with one background event in 1000 days of exposure of $5.6$ tonne fiducial mass \cite{Akerib:2015cja}. The exclusion limit on the scattering cross section for this stringent projection of LZ is expected to be lowered by a factor of $\sim 7 \times 10^{-4}$ compared to the LUX 2013 results \cite{Akerib:2015rjg}. Clearly, this is a rough estimate and a careful analysis of future datasets will be needed.

In this context, we note that LZ projections for anapole DM have also been performed in \cite{Ibarra:2016dlb}\footnote{The paper \cite{Ibarra:2016dlb} studied anapole DM in the context of radiative seesaw models. The parameter space of interest in these models requires values of the anapole moment that are beyond the HL-LHC and LZ projections computed in our paper.}. Our LZ limits are approximately a factor of 2-3 more stringent. For example, for DM with mass $\sim \mathcal{O}(100)$ GeV, we obtain $\mathcal{A}/\mu_N \, \sim \, 4 \times 10^{-7}$ fm, where $\mu_N$ is the nuclear magneton. On the other hand, \cite{Ibarra:2016dlb} adopt the projection limit of $\mathcal{A}/\mu_N \, \sim \, 1 \times 10^{-6}$ fm. Clearly, using the LZ projection of \cite{Ibarra:2016dlb} would only increase the relative importance of our HL-LHC study vis-vis direct detection.

\begin{figure}[!t]
  \centering
\includegraphics[width=.8\textwidth]{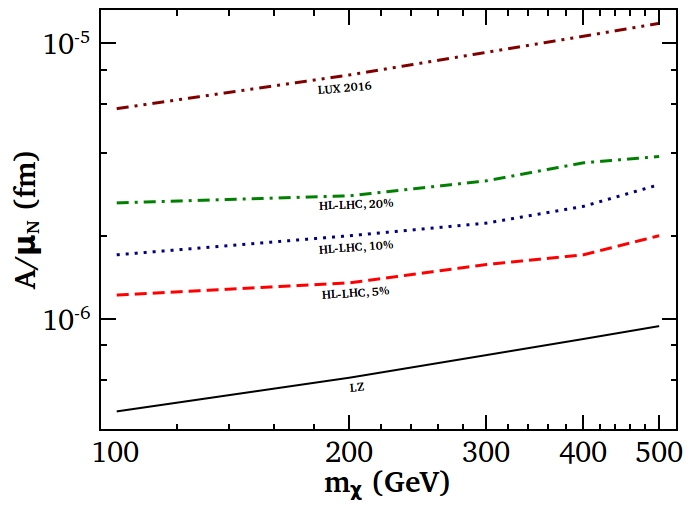}\quad
  \caption{A comparison between the bounds coming from LUX 2016 \cite{Akerib:2016vxi} and projected LZ \cite{Akerib:2015cja} limits on the anapole moment, versus those coming from our HL-LHC study with varying levels of systematic uncertainties.}
  \label{dmcolliderresults}
\end{figure}

The resulting comparative study between the collider reach and the reach from direct detection experiments is shown in Fig.~\ref{dmcolliderresults}. On the horizontal axis, we plot the DM mass in the range 100-500 GeV, while on the vertical axis we plot the anapole moment in units of the nuclear magneton. The LUX 2016 results are shown in the brown dash-dot-dot curve, while the black solid curve shows the projected LZ limit. The green dot-dashed, blue dotted, and red dashed curves show the results obtained from our collider study, with $20\%$, $10\%$, and $5\%$ systematic uncertainties and 3000 fb$^{-1}$ of data at the HL-LHC, respectively.

We can see that there is a vast improvement, almost by an order of magntiude, in the most optimistic projected LZ limit compared to the current LUX limit. This corresponds to the fact that the constraint on the scattering cross section is expected to become stronger by $\sim 100$, although, as we have stressed, this is a rough estimate. On the other hand, the HL-LHC is expected to constrain the anapole moment by a factor of $2-6$ compared to the current LUX results, depending on the level of systematics. It is possible that a study in other channels such as monojet, or a combination of channels, will strengthen the collider results.

\section{Application to a Simplified Model} \label{appsimpmodel}

In this Section, we finally apply our EFT analysis to a specific simplified model. We choose a weakly coupled UV completion in which the DM is a Majorana fermion $\chi$ that couples to an uncolored fermion $f$ (with mass $m_{f}$) and a pair of charged scalars $\widetilde{f}_{L,R}$. At one loop, the DM couples to the photon through a anapole moment interaction. The mass of the charged scalars will be taken to be $\sim 250$ GeV, while the results taken from the EFT will correspond to a cutoff scale $\Lambda \sim 800$ GeV. While the EFT results can be trusted to a first approximation, we note that a detailed collider study of the simplified model will yield more precise constraints.

The  Lagrangian of the model is given by
\begin{equation}
\label{eq:Lint}
  \mathcal{L}_{\text{int}}=\lambda_{L}\widetilde{f}_{L}^{\ast}\overline{\chi}P_{L}f+\lambda_{R}\widetilde{f}_{R}^{\ast}\overline{\chi}P_{R}f+\text{c.c.} \,\,.
\end{equation}

A nonzero mixing angle $\alpha$ is allowed between the scalar mass and chiral eigenstates
\begin{equation}
  \left(\begin{array}{c}
    \widetilde{f}_{1} \\ \widetilde{f}_{2}
    \end{array}\right)
  =\left(
  \begin{array}{cc}
    \cos\alpha & -\sin\alpha \\
    \sin\alpha & \cos\alpha
  \end{array}\right)\left(
  \begin{array}{c}
    \widetilde{f}_{L} \\ \widetilde{f}_{R}
  \end{array}\right)\,.
\end{equation}
The two scalar mass eigenvalues are denoted by  $m_{\widetilde{f}_{1}}$ and $m_{\widetilde{f}_{2}}$. The free parameters of the model are the four masses $m_{\chi}$, $m_{\widetilde{f}_{1}}$, $m_{\widetilde{f}_{2}}$ and $m_{f}$. 

A supersymmetric embedding of this model has been studied in \cite{Fukushima:2014yia, Fukushima:2013efa}. Here, we briefly summarize the dependence of the anapole moment on the model parameters. For a full derivation, we refer to the Appendix of \cite{Sandick:2016zut}.

The relevant Feynman diagrams consist of a triangle loop with either two fermions $f$ or two scalars $\widetilde{f}$, and external legs given by two DM particles and a photon. Let us take the momentum of the incoming and outgoing DM particles to be given by $p$ and $p'$, respectively. The total off-shell scattering amplitude is given by
\begin{equation}
\label{eq:Mmu}
  \mathcal{M}^{\mu}=i\mathcal{A}(q^{2})\overline{u}(p')\left(q^{2}\gamma^{\mu}-\slashed{q}q^{\mu}\right)\gamma^{5}u(p) \,\, ,
\end{equation}
where the momentum transfer is denoted by $q=p'-p$ and the anapole moment by $\mathcal{A}(q^{2})$. The anapole moment $\mathcal{A}(q^{2})$ can be expressed as
\begin{align}
\label{eq:A_alpha}
  \mathcal{A}(q^{2})&=e\left(\left|\lambda_{L}\right|^{2}\cos^{2}\alpha-\left|\lambda_{R}\right|^{2}\sin^{2}\alpha\right)X_{1}(q^{2})\nonumber\\
  &\quad +e\left(\left|\lambda_{L}\right|^{2}\sin^{2}\alpha-\left|\lambda_{R}\right|^{2}\cos^{2}\alpha\right)X_{2}(q^{2}) \,\,,
\end{align}
where $X_{1,2}$ is the result of three-point loop integrals. The derivation of these equations, along with the full form of $X_{i}$, is given in \cite{Sandick:2016zut}.  

In the limit $|q^{2}|\ll m_{f}^{2}$ and $|q^{2}|\ll m_{\widetilde{f}_{i}}^{2}$, the $X_{i}$ reduce to a simple expression,
\begin{align}
X_{i} [{q^{2}=0}] \,\, \longrightarrow \,\, \frac{1}{96\pi^{2}m_{\chi}^{2}}\left[\frac{3\mu_{i}-3\delta+1}{\sqrt{\Delta_{i}}}
 {\rm arctanh} \left(\frac{\sqrt{\Delta_i}}{\mu_{i}+\delta-1}\right)-\frac{3}{2}\log\left(\frac{\mu_{i}}{\delta}\right)\right],
 \label{exprXi}
\end{align}
where $\Delta_{i}=(\mu_{i}-\delta-1)^{2}-4\delta$, $\delta = m_f^2/m_\chi^2$, and $\mu_{1} =\frac{m^{2}_{\widetilde{f}_{1}}}{m_{\chi}^{2}}\,,  \, \mu_{2}=\frac{m^{2}_{\widetilde{f}_{2}}}{m_{\chi}^{2}}\,, \,\delta=\frac{m_{f}^{2}}{m_{\chi}^{2}}$. This limit applies to DM direct detection for $f=\mu, \tau$. For very heavy mediators $\mu_{i}\gg 1$,  $X_{i}$ vanishes as $\mu_{i}^{-1}\log\mu_{i}$. If the mass difference between $\widetilde{f}$ and the DM is small, the value of $X_{i}$ will increase; in the limit $(\mu_{i}-1)\sim\delta\ll 1$,
\begin{equation}
\label{eq:Xi}
  X_{i}\sim\frac{1}{96\pi^{2}m_{\chi}^{2}}\left[\frac{\pi}{\sqrt{\delta}}-\frac{3}{2}\log\frac{1}{\delta}\right].
\end{equation}
For $f=\mu$ and $\tau$, this model has a sizable anapole moment, which can be detected in direct detection experiments. We now present the limits from various experiments shown in Fig.~\ref{dmcolliderresults} in the parameter space of this class of models.

\begin{figure}[!t]
  \centering
\includegraphics[width=.8\textwidth]{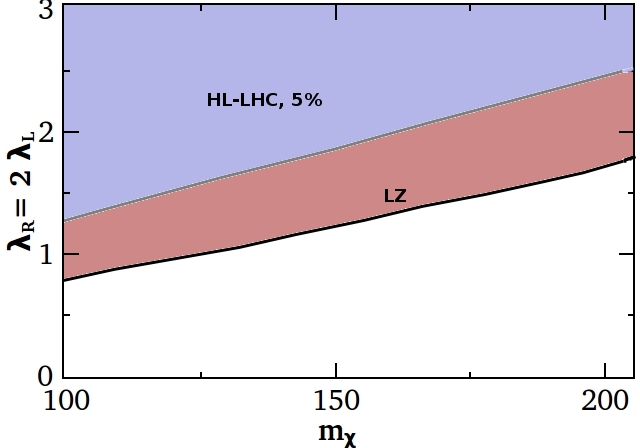}\quad
  \caption{Constraints on the $(\lambda, m_{\chi})$ plane: We show the constraints on the $\lambda_R = 2 \lambda_L$ versus the DM mass $m_{\chi}$ plane, keeping $\alpha = \pi/4$ and $\mu_{1} \, = \, m^{2}_{\widetilde{f}_{1}}/m_{\chi}^{2} \, = \, 1.44$. The bluish grey region shows the part of parameter space that will be constrained by mono-$Z$ searches at the HL-LHC, assuming 3000 fb$^{-1}$ of data and $5\%$ systematic uncertainties. The red region is the region that will be constrained by LZ, assuming the most optimistic performance with one background event in 1000 days of exposure of $5.6$ tonne fiducial mass. Current LUX limits and direct collider searches for the mediators $\widetilde{f}_{1}$ are not able to constrain this part of parameter space.}
  \label{lambda_versus_mass}
\end{figure}

In Fig.~\ref{lambda_versus_mass}, we first plot the constraints on the $(\lambda, m_{\chi})$ plane. On the vertical axis, we plot $\lambda_R = 2 \lambda_L$, while on the horizontal axis we plot $m_\chi$ in the range 100-200 GeV. We keep the mixing angle fixed at $\alpha = \pi/4$. The mass of the lightest scalar mediator $\widetilde{f}_{1}$ is kept at
\be
\mu_{1} \, = \, \frac{m^{2}_{\widetilde{f}_{1}}}{m_{\chi}^{2}} \, = \, 1.44 \,\,\,\,\,\,.
\ee
In the region of parameter space plotted, the only constraints come from our HL-LHC study and LZ projections. Current LUX constraints on the anapole moment are too weak to show up in this region, while direct search constraints for the uncolored mediator $\widetilde{f}_{1}$ lie below 100 GeV. The bluish grey region shows the part of parameter space that will be constrained by mono-$Z$ searches at the HL-LHC, assuming 3000 fb$^{-1}$ of data and $5\%$ systematic uncertainties. The red region is the region that will be constrained by LZ, assuming the most optimistic performance with one background event in 1000 days of exposure of $5.6$ tonne fiducial mass.

We note that lower values of $\mu_1$, corresponding to a greater degree of compression between the lightest scalar mediator and the DM, have larger values of the anapole moment from Eq.~(\ref{exprXi}). These regions are constrained both by our HL-LHC study and by LZ projections, indeed to a greater extent than is the case in Fig.~\ref{lambda_versus_mass}. On the other hand, larger values of $\mu_1$ are less suited to our collider search strategy as well as DM direct detection, due to a reduced value of the anapole moment. These regions start becoming constrained by collider searches for the mediators $\widetilde{f}$ themselves, as the mass gap between them and the DM increases leading to collider signals with large missing energy.

We show the range of DM masses between 100-200 GeV. Below $m_{\chi} = 100$ GeV, there are LEP constraints on the uncolored scalar mediators. Above 200 GeV, the anapole moment becomes smaller and the constraints become less stringent.

Finally, changing the value of $\alpha$ also changes the anapole moment. We turn to this dependence next.

\begin{figure}[!t]
  \centering
\includegraphics[width=.8\textwidth]{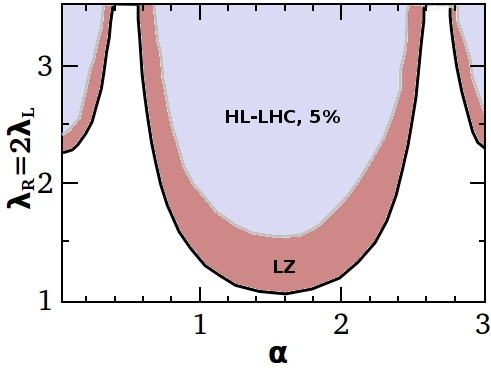}\quad
  \caption{Constraints on the $(\lambda, \alpha)$ plane: We show constraints on the $\lambda_R = 2 \lambda_L$ versus $\alpha$ plane, keeping $m_{\chi} = 200$ GeV and $\mu_{1} \, = \, m^{2}_{\widetilde{f}_{1}}/m_{\chi}^{2} \, = \, 1.44$. The bluish grey region shows the part of parameter space that will be constrained by mono-$Z$ searches at the HL-LHC, assuming 3000 fb$^{-1}$ of data and $5\%$ systematic uncertainties. The red region is the region that will be constrained by LZ, assuming the most optimistic performance with one background event in 1000 days of exposure of $5.6$ tonne fiducial mass. Current LUX limits and direct collider searches for the mediators $\widetilde{f}_{1}$ are not able to constrain this part of parameter space.}
  \label{lambda_versus_alpha}
\end{figure}

In Fig.~\ref{lambda_versus_alpha}, we display the constraints on the  $(\lambda, \alpha)$ plane. We keep the mass of DM and the lightest scalar mediator fixed at  $m_{\chi} = 200$ GeV and $\mu_{1} \, = \, m^{2}_{\widetilde{f}_{1}}/m_{\chi}^{2} \, = \, 1.44$, respectively. The color scheme is the same as in the previous figure. We see the presence of ``blind regions" - regions near $\alpha = \pi/8, \, 7 \pi/8$ - where the anapole moment becomes highly suppressed, from Eq.~(\ref{eq:A_alpha}). These regions are difficult to probe using any method that relies on the photon coupling; in \cite{Sandick:2016zut}, however, it was shown that indirect detection can constrain these regions. The effect of changing either $\mu_1$ (and hence the light mediator mass) or the mass of the DM has been discussed before, and applies to this figure as well. Lowering $\mu_1$ leads to larger values of the anapole moment and stronger constraints on the  $(\lambda, \alpha)$ plane, and the ``blind regions" get sharpened to values very close to $\alpha = \pi/8, \, 7 \pi/8$. Increasing $m_{\chi}$ weakens the collider and direct detection constraints.

\begin{figure}[!t]
  \centering
\includegraphics[width=.8\textwidth]{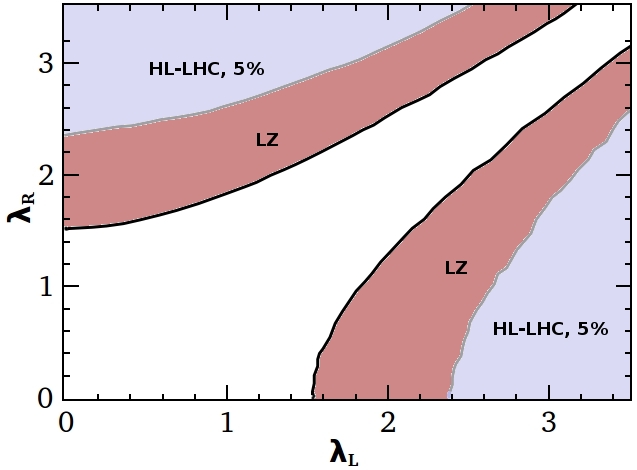}\quad
  \caption{Constraints on the $(\lambda_R, \lambda_L)$ plane: We show constraints on the $\lambda_R$ versus $\lambda_L$ plane, keeping $m_{\chi} = 200$ GeV, $\mu_{1} \, = \, m^{2}_{\widetilde{f}_{1}}/m_{\chi}^{2} \, = \, 1.44$, and $\alpha = \pi/4$. The bluish grey region shows the part of parameter space that will be constrained by mono-$Z$ searches at the HL-LHC, assuming 3000 fb$^{-1}$ of data and $5\%$ systematic uncertainties. The red region is the region that will be constrained by LZ, assuming the most optimistic performance with one background event in 1000 days of exposure of $5.6$ tonne fiducial mass. Current LUX limits and direct collider searches for the mediators $\widetilde{f}_{1}$ are not able to constrain this part of parameter space.}
  \label{lambda_versus_lambda}
\end{figure}

Finally, in Fig.~\ref{lambda_versus_lambda}, we plot our results in the $(\lambda_R, \lambda_L)$ plane, keeping $m_{\chi} = 200$ GeV, $\alpha = \pi/4$, and $\mu_1 = 1.44$. The color scheme is the same as in the previous figures. The corridor around the region where $\lambda_R \sim \lambda_L$ constitutes a ``blind region" where the anapole moment is attenuated for $\alpha = \pi/4$. These regions are difficult to probe for methods that rely on the photon coupling. Parts of this region can be explored by indirect detection.

\section{Conclusions}

In this paper, we have explored the HL-LHC detection prospects for DM that couples to the Standard Model through higher electromagnetic moments, particularly the anapole moment. The study is conducted at the level of EFT, with the aim of calculating the reach in the cutoff scale $\Lambda$. We have conducted our study in the mono-$Z$ channel, taking into account varying levels of systematic uncertainties. We carefully choose kinematic variables that can discriminate between signal and SM background, and select cuts using the Bayesian optimization method implemented in the Python algorithm \HyperOpt . A BDT is then used to classify events into signal and background classes.

The results of our collider study are shown in Fig.~\ref{reachLambda}. We see that for a very small systematic uncertainty of 1\%, the $5\sigma$ reach in $\Lambda$ is above 1.1 TeV for DM masses in the range 100-500 GeV. Assuming a larger systematic uncertainty of 5\% , a 100 GeV DM can still be discovered for $\Lambda \approx$ 1 TeV. From 5\% to 10\% and from 10\% to 20\% systematics, the discovery reach in $\Lambda$ decreases by 200 GeV for a given DM mass.

The  $5\sigma$ reach in the cutoff scale $\Lambda$ obtained from our collider study can be mapped on to a reach in the value of the anapole moment $\mathcal{A}$ through Eq.~(\ref{anapoleEFT}). The resulting comparative study between the collider reach and the reach from direct detection experiments is shown in Fig.~\ref{dmcolliderresults}. The LUX 2016 results are shown in the brown dash-dot-dot curve, while the black solid curve shows the projected LZ limit. The green dot-dashed, blue dotted, and red dashed curves show the results obtained from our collider study, with $20\%$, $10\%$, and $5\%$ systematic uncertainties and 3000 fb$^{-1}$ of data at the HL-LHC, respectively.

Finally, the EFT analysis is applied to a specific simplified model. We choose a weakly coupled UV completion in which the DM is a Majorana fermion $\chi$ that couples to an uncolored fermion $f$ (with mass $m_{f}$) and a pair of charged scalars $\widetilde{f}_{L,R}$. At one loop, the DM couples to the photon through a anapole moment interaction. The HL-LHC constraints on the parameter space of this class of models is presnted in Fig.~\ref{lambda_versus_mass} - Fig.~\ref{lambda_versus_lambda}. These constraints are juxtaposed with projected LZ constraints on the parameter space, assuming the most optimistic performance with one background event in 1000 days of exposure of $5.6$ tonne fiducial mass.

\section{Acknowledgements}
A. Alves acknowledges financial support from the Brazilian
agencies CNPq, under the process 307098/2014-1, and Funda\c{c}\~ao de Amparo \`a Pesquisa do Estado de S\~ao Paulo (FAPESP), under the process 2013/22079-8. A. C. O. Santos acknowledges Coordena\c c\~ao de Aperfei\c coamento de Pessoal de N\'ivel Superior (Capes-PDSE-88881.135139/2016-01).

\end{document}